\documentclass[aps,superscriptaddress,showpacs,onecolumn]{revtex4}

\usepackage{graphics}
\usepackage{dcolumn}
\usepackage{epsfig}
\usepackage{amsmath}
\usepackage{mathrsfs}
\usepackage{amsfonts}
\usepackage{amssymb}
\begin{document}

\title{\textbf{MagGene}: A genetic evolution program for magnetic structure prediction}

\author{Fawei Zheng}
\email{zheng_fawei@iapcm.ac.cn}
\affiliation{LCP, Institute of Applied Physics and Computational Mathematics, Beijing, China}
\author{Ping Zhang}
\email{zhang_ping@iapcm.ac.cn}
\affiliation{LCP, Institute of Applied Physics and Computational Mathematics, Beijing, China}
\affiliation{School of Physics and Physical Engineering, Qufu Normal University, Qufu, China}
\affiliation{Beijing Computational Science Research Center, Beijing 100084, China}

\date{\today}

\begin{abstract}
We have developed a software \texttt{MagGene} to predict magnetic structures by using genetic algorithm. Starting from an atom structure, \texttt{MagGene} repeatedly generates new magnetic structures and calls first-principles calculation engine to get the most stable structure. This software is applicable to both collinear and noncollinear systems. It is particularly convenient for predicting the magnetic structures of atomic systems with strong spin-orbit couplings and/or strong spin frustrations.
\end{abstract}


\maketitle

{\bf PROGRAM SUMMARY}

\begin{small}
\noindent\\[-0.25cm]
{\em Program Title:} \texttt{MagGene}                \\ \\[-0.25cm]
{\em Catalogue identifier:}                                    \\ \\[-0.25cm]
{\em Programming language:} Fortran 90\                        \\ \\[-0.25cm]
{\em Computer:} any computer architecture                      \\ \\[-0.25cm]
{\em RAM: } system dependent, at least 10 MB                     \\ \\[-0.25cm]
{\em Program obtainable from: }                               \\ \\[-0.25cm]
{\em Number of processors used:} 1                             \\ \\[-0.25cm]
{\em CPC Library subroutines used:}    None                    \\ \\[-0.25cm]
{\em Operating system:} Linux, Windows, Mac                     \\ \\[-0.25cm]
{\em External routines/libraries:} None                \\ \\[-0.25cm]
{\em Nature of problem:}
In complex atomic systems , such as systems with strong spin-orbit coupling and/or strong spin-frustrations, the associated magnetic structures could also be quite complex. The traditional methods, such as analysis based on crystal symmetries and simulated anealing based on a special model Hamiltonian, are inefficent. Then the electronic structures, spin wave dispersions, and other properties based on correct magnetic structures could not be obtained accurately. Therefore, an efficient method to predict magnetic structures is required.
\\ \\[-0.25cm]
{\em Solution method:}
The magnetic structures can be predicted efficiently by using genetic algorithm.
\\ \\[-0.25cm]
{\em Running time:} system dependent, from a few minutes to several weeks \\ \\[-0.25cm]
{\em Unusual features of the program:} The magnetic structures can be predicted for both collinear and noncollinear atomic systems by using genetic evolution algorithm. The total magnetic moment can be fixed. \\ \\[-0.25cm]
{\em References:} \\

\end{small}

\section{Introduction}
A simple magnetic material usually has a relatively simple magnetic structure. It can be ferromagnetic, antiferromagnetic, or ferrimagnetic. The magnetic structure can be measured by neutron scattering experiment. But, the experiment may be failed to apply to many systems. For example, ultra thin two-dimensional (2D) systems, such as FeSe single layer\cite{FeSeLayer}, have too limited scattering cross-sections to perform neutron scattering experiment. Another example is the materials whose qualified crystals are not yet obtained, hence the magnetic structures are impossible to be determined by neutron scattering experiment. Then the theoretical prediction becomes important. Normally we have two ways to theoretically determine the magnetic structures. One way is guessing-calculating. We firstly guess some possible magnetic structures based on the nature of magnetic atoms and the crystal symmetries. Then, we calculate the energy for each structure by using total energy calculation method, such as density functional theory (DFT) \cite{DFT1,DFT2} and Hartree-Fock\cite{HF1,HF2,HF3}.  If we have exhausted all possible magnetic structures, then the most stable structure is the one we required. The other theoretical way is based on model Hamiltonian. Firstly, we suppose the system has a special kind of model Hamiltonian. Then we fit the model parameters in a series of total energy calculations or experiment observations. After the Hamiltonian is obtained, the magnetic structure can be predicted by theoretical analysis or simulated annealing calculations. Unfortunately, both the two methods may be failure in complex systems. In such a complex system, the real magnetic structure may also be quite complex, therefore it is hard to guess in advance and the guessing-calculating method fails. The model Hamiltonian for this system is probably also very complex. Then the task to get an accurate model Hamiltonian is not easy. Then the second method fails too. 

The complexity of magnetic structure prediction is attributable to the vast amount of possible structures. Actually, there are many other problems share the same fundamental feature. One is deep leaning, which is one of the most remarkable machine learning techniques. It has achived great successes in many applications such as image analysis, speech recognition, and text understanding. There are many layers in a deep neural network, and each layer may have hundreds of neurons. The total number of tunable parameters is typically more than ten thousand. Besides, there are hyperparameters which determines the structure of a neural network. Then the number of possible configurations is virtually infinite.  In the past few years, the genetic algorithm has been found to be quite useful in deep learning.  The genetic algorithm can be used to train deep convolutional neural networks\cite{DL1} and deep reinforcement learning neural networks\cite{DL2}. It trained a neural network to become a super-strong StarCraft game player\cite{DL3}. Another example, which is more closely related to magnetic structure prediction, is the prediction of crystal structure. The traditional way to get crystal structure is based on experiment, which needs certain amount of synthesized crystal. But, many crystals are hard to synthesize or only exist at extrem conditions. Then the theoretical prediction is necessary. The prediction is only on the knowlege of chemical compositions, and is extremely difficult involving a huge number of energy minima on energy surface.  Among variety theoretical crystal prediction methods, the genetic algorithm\cite{CS1,CS2,CS3,CS4,CS5,CS6,CS7,CS8} and particle swarm optimization algorithm\cite{CS9} are remarkable. Thousands of new materials have been predicted by using them. Following these predictions, many of them have also been successfully experimentally synthesized. 

Considering the successful applications of genetic algorithm in deep learning and crystal structure prediction, we believe that it could also apply to magnetic structure prediction. In the past few years, we have been continuously working on this problem, and developed a python program\cite{scr}. The program can predict the magnetic structures in collinear magnetic systems. It is the embryonic form of \texttt{MagGene}. Besides this code, the latest version (v.10.2) of crystal structure prediction code \texttt{Uspex}\cite{CS3} can also predict collinear magnetic structures of materials.  In this paper, we introduce a new version of \texttt{MagGene} code. The code is written in fortran, based on genetic algorithm. It is applicable to both collinear and noncollinear magnetic systems. A genetic algorithm is a search or optimization algorithm inspired by Charles Darwin's theory of natural evolution. The algorithm repeatedly modifies a population which is defined as a set of indivuduals. Each individual here is a solution to the problem you want to solve, and it is a magnetic structure in the present problem. At each step, the genetic algorithm randomly selects structures to be parents and produces children structures to form the population in the next generation. Using a certain fitness function, the genetic algorithm will continuously throw the weak structures and keep the strong ones. And finally, the population would evolve to optimal structures. \texttt{MagGene} uses total energy, calculated by \texttt{Vasp} code\cite{VASP}, as the fitness function. Following the genetic evolvement, \texttt{MagGene} will find the best magnetic structures which have the lowest total energies. 

During the genetic evolvement, many structures may evolve to one single meta-stable structure and be trapped there, especially when the system is complex and the evolvement has been running for long time. Then the population losts diversity which plays an important role in genetic algorithm. Losing diversity, the genetic algorithm may be trapped in a meta-stable state, and can not continue searching for the real ground state. In order to overcome this difficulty, we have introduced two techniques in \texttt{MagGene}. The first technique is simply introducing random structures in each generation. The other technique is to throw duplicate structures. In the case of collinear magnetic system, we compare each new generated structure and delete the duplicated ones. For the case of noncollinear magnetic system, we calculate the similarity ($s$) between every two magnetic structures. For example, the similarity between the $i$-th and $j$-th magnetic structures is $s_{i,j}=\sqrt{\sum_{n=1...N}^{t=x,y,z}(M_{i}^{n,t}-M_{j}^{n,t})^2/N}$. Here, the $M_{i}^{n,t}$ is the $t$ component of magnetic moment for the $n$-th magnetic atom in $i$-th configuration. If the similarity between two structures belows a critic value that is controled by keyword {\bf similarity\_criteria}, then \texttt{MagGene} will throw one of the structures depending on their relative stabilities.  

Besides the diversity problem, we find that an unconverged calculation may also cause problem. The magnetic systems are more complex than nonmagnetic systems. Hence, a magnetic system is harder to converge. Sometimes, it is very tricky in choosing parameters to make the calculation converge. The unconverged calculation stops with an inaccurate total energy. Fortunately, our experiences show that the true ground state usually converges much faster than the states with high energies. Due to their high energies, the unconverged state may have no harm to the final results of genetic evolvement. But their large error bars make them have a certain possibility to provide a fake ground state. The possibility increases with increasing the genetic evolvement time, and would finally becomes fatel. To avoid this predicament, \texttt{MagGene} set a very high pseudo energy for each unconverged structure. Therefore, these structures would have disadvantage comparing with the true ground state, and the fake ground state is avoided. Based on genetic algorithm associated with  techniques to enhance diversity and emperical method to eliminate fake ground state, \texttt{MagGene} can search for the ground magnetic structure efficiently and reliably.

This paper is organized as follows. The introductions of workflow in \texttt{MagGene} are presented in Section II; the detailed descriptions of control parameters are shown in Section III; three groups of examples are shown in Section IV, followed by the conclusion in Section V.

\section{Brief description of the code}

After decompressing the {\it MagGene.tar.gz} file, we get a folder named as {\it MagGene}. Inside, there are {\it makefile} and {\it Readme} files and three sub-folders. They are {\it src}, {\it doc} and {\it examples}. Folder {\it src} contains all the Fortran90 source codes, folder {\it examples} contains six examples, and folder {\it doc} contains flowchart and a text file describing the keywords used in \texttt{MagGene}. The main input file is {\it input.dat}, which contains all the control parameters. Its detailed descriptions will be shown in the next section. 

\begin{figure}
\includegraphics[width=0.5\columnwidth]{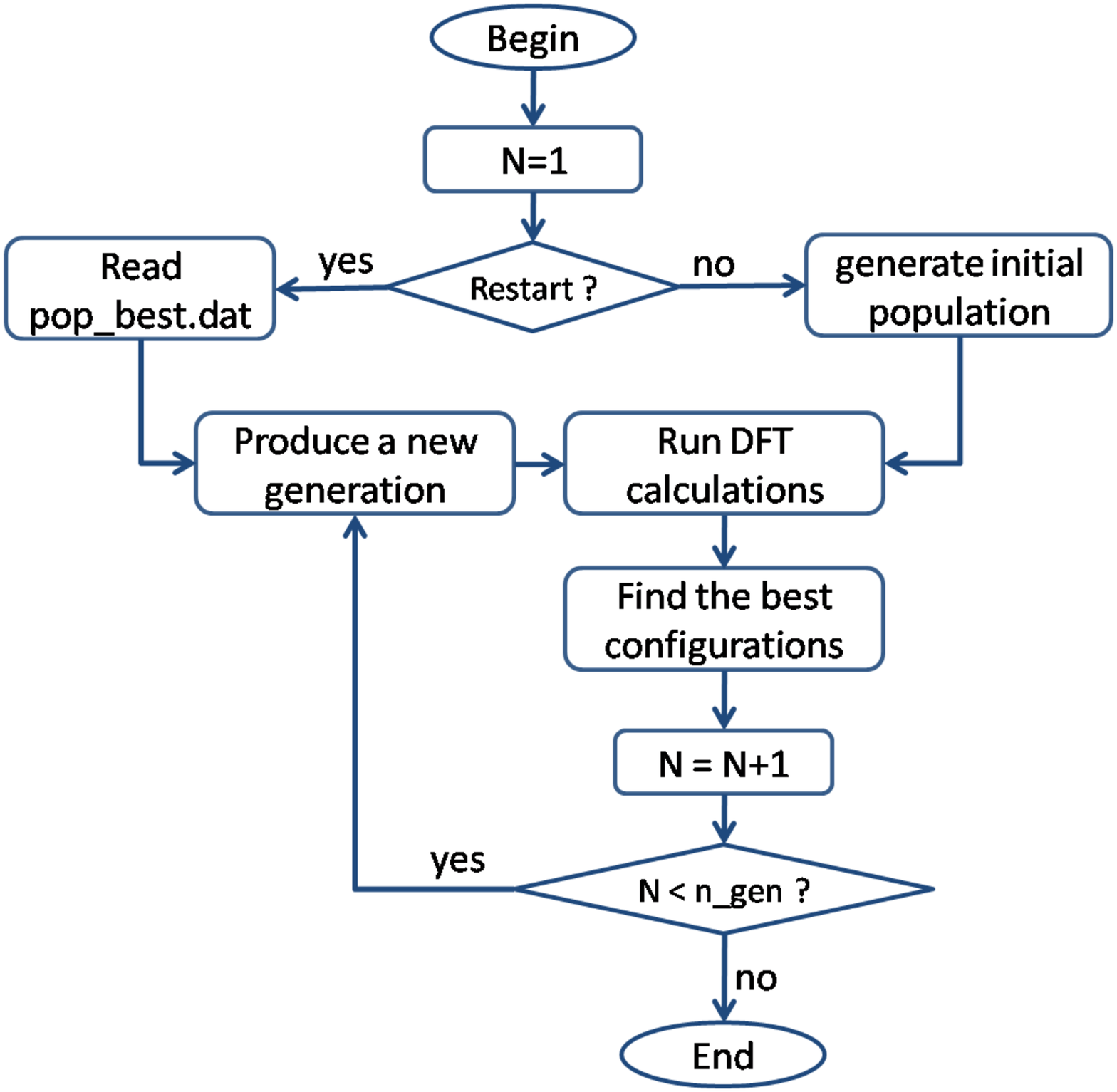}
\caption{\label{Fig1} Flow chart of magnetic structure prediction process in \texttt{MagGene}.}
\end{figure}

The data structures and treatments for collinear and noncollinear magnetic materials are quite different in genetic algorithm. In the collinear systems, the magnetic moment of each magnetic atom can be described by up or down spin. Only one real number is needed for a magnetic atom. Then the crossover and mutation of magnetic structures are described by the combination and distortion of a group of one dimensional real arrays. While, in the case of noncollinear systems, the magnetic moment of each magnetic atom can point to any direction in three dimensional spherical surface. We have to describe a magnetic structure by a two-dimensional real array. The crossover and mutation of magnetic structures are more complex. Therefore, in the code, the collinear and noncollinear parts are written separately. In calculaion, the code will firstly read keyword {\bf noncollinear}. If the value is {\bf 1} ({\bf 0}), then \texttt{MagGene} will call the noncollinear (collinear) part of the code. Although the data structures and treatments are different, the main calculations for collinear and noncollinear systems follow the same flowchart as shown in Fig. \ref{Fig1}. The first generation of population is generated randomly or based on previously stored magnetic structures. The later case is called restart mode, which can be activated by {\bf restart =1}. In the restart mode, \texttt{MagGene} will read magnetic structures from {\it pop\_best.dat} file. It is generated by a previous \texttt{MagGene} calculation. We can also edit {\it pop\_best.dat} to take advantage of our prior knowlege of the system. After having generated the first generation of population, \texttt{MagGene} will perform DFT calculations by using shell scripts written in {\it submit.sh}, which contains the command calling \texttt{Vasp}. After the total energies of all the structures are calculated, MagGene decides which structures will survive. Then, the code will generate the next generation of population, and continue to perform the DFT calculations. The calculation continues until reaches the maximum cycle number defined by {\bf n\_gen}. 

In the 'Produce a new generation' step of FIg. \ref{Fig1}, \texttt{MagGene} performs the crossover, mutation, as well as random structure generation. The number of generated structures in the new generation is controlled by {\bf population}, {\bf n\_mutation}, and {\bf n\_random}. The code will randomly choose two structures that survived in previous calculations, and combine these two parent structures to produce a child structure, which is the crossover structure. The total number of crossover structure is {\bf population}-{\bf n\_mutation}-{\bf n\_random}. We carry out the mutation process by distorting previously survived structures. The extent of distortion is controlled by {\bf m\_mutation} and {\bf delta\_mutation} for collinear and noncollinear systems respectively. The total number of mutation structures is {\bf n\_mutation}. \texttt{MagGene} allows each generation to contain {\bf n\_random} number of random structures. It incrases the diversity of population, and avoides the system to be trapped in meta-stable structures. As shown in the last section, the other important way to incrase diversity is to throw the similar contructures. Before DFT calculations in each genetic cycle, \texttt{MagGene} checkes the similarity of the new generated structures and the previously calculated structures. Then throw the similar structures. The same procedure will perform again after DFT calculations, since the magnetic structures may be altered slightly during DFT calculation especially for the noncollinear systems.

Besides increasing the diversity of population, during our testing, we found an emperical method to enhance the quality of genetic evolution. There is a possibility that an unconverged state becomes a fake ground state and damages the quality of genetic evolution.
To avoid this predicament, \texttt{MagGene} checkes the convergence of each DFT calculation, and adds a large value to the total energy of each unconverged state. Therefore, the final best structure is always the converged true ground state. This function can be activated by setting {\bf dft\_converge=1}. We find that this emperical method is especially significant for complex systems.

\texttt{MagGene} produces many different output files to show the important informations. In the main folder, the code produces {\it pop.dat}, {\it pop\_best.dat}, {\it energy.dat}, {\it diversity\_matrix.dat}, and {\it best\_structure\_{\it n}.xsf} files. All the structures that have been calculated are stored in {\it pop.dat} file, and the best strutures are stored in {\it pop\_best.dat} file. The differences between each two best structures form a matrix stored in {\it diversity\_matrix.dat} to show the diversity of the best structures. The total energies of the best structures are stored in {\it energy.dat} file, which can be plotted to show the convergence of genetic evolution. The magnetic structures of the best structures are also written in a series of {\it .xsf} files which can be conveniently shown by using \texttt{Xcrysden}\cite{Xcrysden}. Besides these files, there are also several sub-folders named as {\it gen\_n} under the main folder. Sub-folder {\it gen\_n}' contains all the {\it INCAR}, {\it OSZICAR}, and {\it OUTCAR} files for {\it n}-th generation. So that we can check the informations of each \texttt{Vasp} calculation. In each generation, the total energies and structures are also written in a {\it energy\_gen\_n.dat} file and a group of {\it .xsf} files in the associated sub-folder for convenient inspection and visualization. 

\section{The input.dat file}
All the control parameters of \texttt{MagGene} and most of the magnetic atom informations can be described by keywords. They are provided in {\it input.dat} file, and listed in no particular order. Case is ignored, so that {\bf noncollinear} is the same as {\bf Noncollinear} and {\bf NONCOLLINEAR}. Characters after "!" or "\#" are treated as comments. Most of the keywords have default values.
The keywords are described in detail as follows:

\begin{itemize}
\item{ {\bf noncollinear} }
    \\ {\it Default value} : 0
    \\ {\it Value type} : integer
    \\[-0.25cm]
    \\The keyword {\bf 'noncollinear'} determine whether the material is a noncollinear magnetic system. The value of {\bf 'noncollinear'} has two available options, they are:
    \\[-0.25cm]
    \\0 :  collinear magnetic system.
    \\[-0.25cm]
    \\1 : noncollinear magnetic system.

\item{{\bf restart}}
\\{\it Default value} : 0
\\ {\it Value type} : integer
\\[-0.25cm]
\\ Available options are:
 \\           0 : run genetic agrithm from randomly generated initial structures
 \\           1 : read pop\_best.dat file, and run genetic agrithm from previously stored best structures.    
    
\item{{\bf m\_atom\_list} }
\\{\it Default value} : no default values
\\{\it Value type} : integer list
\\[-0.25cm]
\\ The keyword {\bf m\_atom\_list} shows the atoms that have non-zero magnetic moment. For example {\bf m\_atom\_list} = 2 3 6 means that the second, third and sixth atoms are magnetic.

\item{{\bf generation} }
\\{\it Default value} : 1
\\{\it Value type} : integer
\\[-0.25cm]
\\Define the number of generations to be calculated..

\item{{\bf population} }
\\{\it Default value} : 1
\\ {\it Value type} : integer
\\[-0.25cm]
\\Define the total number of population, i.e. the total number of different structures in one generation.

\item{{\bf similarity\_criteria} }
\\{\it Default value} : 1 (collinear); 0.1 (noncollinear)
\\ {\it Value type} : integer(collinear); real(noncollinear)
\\[-0.25cm]
\\Define the criteria used to delete similar structures. In collinear systems, the similarity\_criteria is the total number of magnetic atoms that have different magnetic directions (up and down) in two magnetic structures. In noncollinear systems, the similarity\_criteria is defined by $s_{i,j}=\sqrt{\sum_{n=1...N}^{t=x,y,z}(M_{i}^{n,t}-M_{j}^{n,t})^2/N}$.

Used in the case of noncollinear = 1.

\item{{\bf pop\_best}}
\\{\it Default value} : 1
\\ {\it Value type} : integer
\\[-0.25cm]
\\ Define the number of the best structures to be saved in each generation.

\item{{\bf n\_random}}
\\{\it Default value} : 0
\\ {\it Value type} : integer
\\[-0.25cm]
\\ The number of random structures added to each generation.

\item{{\bf n\_mutation}}
\\{\it Default value} : 0
\\ {\it Value type} : integer
\\[-0.25cm]
\\  The number of randomly distorched structures added to each generation. Each structure is generated by distorting one of the best structures.

\item{{\bf m\_mutation}}
\\{\it Default value} : 0
\\ {\it Value type} : integer
\\[-0.25cm]
\\ Used in the case of n\_mutation $>$ 0 and noncollinear = 0 to generate randomly distorched structures. It defines the number of spin flippings in each structure.

\item{{\bf delta\_mutation}}
\\{\it Default value} : 0.0
\\ {\it Value type} : real
\\[-0.25cm]
\\  Used in the case of n\_mutation $>$ 0 and noncollinear = 1 to generate randomly distorched structures. It defines the maximum value of spin rotation.  delta\_mutation = max($dm_x^2+dm_y^2+dm_z^2$).

\item{{\bf norm\_mag}}
\\{\it Default value} : 1.0
\\ {\it Value type} : real list
\\[-0.25cm]
\\ The norm of each atomic magnetic momentum. The list should have the same number of elements in m\_atom\_list.

\item{{\bf fixm}}
\\{\it Default value} : 1.0d7 (collinear); 1.0d7, 1.0d7, 1.0d7 (noncollinear)
\\ {\it Value type} : real (collinear); 3*real (noncollinear)
\\[-0.25cm]
\\ If fixm is provided and the value is smaller than 1.0d6, then the total magnetic moment of the system is constrained to fixm. As a special case, the code will perform antiferromagnetic order searching if  fixm = 0.0 (collinear) or fixm = 0.0 0.0 0.0 (noncollinear).

\item{{\bf afm}}
\\{\it Default value} : 0
\\ {\it Value type} : integer
\\[-0.25cm]
\\ Available options are:
 \\           0 : no effect
 \\           1 : reset fixm =0.0 (collinear) or fixm = 0.0 0.0 0.0 (noncollinear)

\item{{\bf dft\_converge}}
\\{\it Default value} : 0
\\ {\it Value type} : integer
\\[-0.25cm]
\\ Available options are:
 \\           0 : no effect
 \\           1 : increase the energy by a large value when the DFT calculation is not converged after reaching the largest self-consistent cycle number.

\end{itemize}

\section{Examples}
In the following context we will illustrate the capabilities of \texttt{MagGene} by three examples: (i) bulk and single layer FeSe, which are iron-based superconducting materials with two different collinear magnetic structures; (ii) CrI$_3$ single layers with different lattice parameters, which are two-dimensional magnetic materials with spin-orbit couplings; (iii) UO$_2$ bulk, which is a nuclear material with a very peculiar noncollinear magnetic structure. All DFT calculations are performed by using \texttt{Vasp} package\cite{VASP}. The core electrons are described by PAW method\cite{PAW1,PAW2,PAW3}. All the other DFT settings and genetic algorithm parameters will be shown separately in each case. The convergences of these DFT settings have already been tested. The atomic structures are relaxed in advance until the force on each atom is smaller than 0.005 eV/\AA.  The relaxing of atomic structures during genetic evolvement is important for some systems, especially when the magnetic structures have strong correlations with the associated atomic structures. While the relaxing would significantly increase the time consuming. So, it is a problem to balance the accuracy and efficiency. In the present examples, different magnetic structures only lead to negligible atomic structure distortions. Therefore, only the electron self-consistent calculation is needed in genetic algorithm procedures.

\subsection{FeSe Bulk and Single Layer}
FeSe is one of the most simple iron-based superconductor. It has a quasi-two dimensional structure without any intercalated atoms. The superconducting temperature is found to be around 8 K\cite{FeSeBulk}. In first-principles calculations, its most stable magnetic structure is  pair-checkerboard antiferromagnetic\cite{FeSeGong}. A single layer of FeSe can be fabricated on the (100) surface of SrTiO$_3$\cite{FeSeLayer}. Compare with FeSe bulk, the single layer form of FeSe has an astonishingly high superconducting temperature (65 K).  The system has many different complex possible magnetic structures, which may strongly modulate the electronic structures\cite{FeSeLu,FeSeBazhirov,FeSeZheng1,FeSeZheng2}. In first-principles calculations, the most stable magnetic structure of FeSe single layer is collinear antiferromagnetic\cite{FeSeLu}.

\begin{figure}
  \includegraphics[width=0.5\columnwidth]{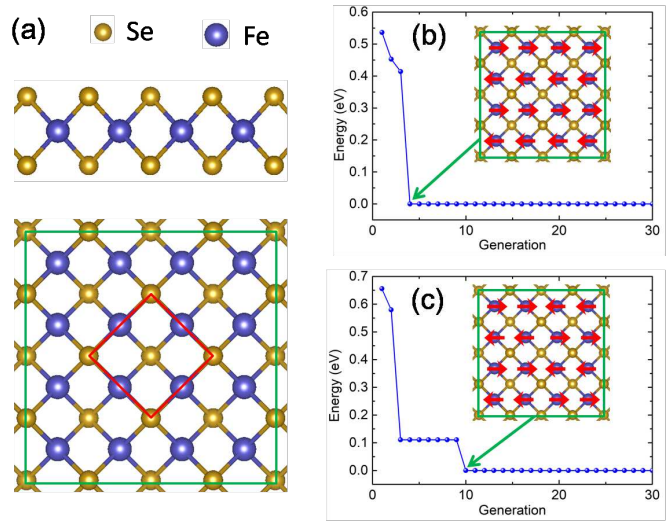}
  \caption{
  \label{Fig2}
  (a) Atomic structure of one FeSe layer with its in-plane supercell (green rectangle) and primary cell (red rectangle). Panels (b) and (c) shows the {\bf MagGene} calculation results for single layer and bulk form of FeSe respectively. The predicted final structures are also shown in the inner panel.}
\end{figure}

As the first example, we performed the genetic evolvement calculations to find the most stable magnetic structures of FeSe bulk and single layer. The atomic structure of one layer of FeSe is shown in Fig. \ref{Fig2}(a). In which, the violet and yellow balls are Fe and Se atoms respectively. The red rectangle is the primitive cell. The in-plane and out-plane lattice parameters of FeSe bulk are 3.77 \AA\, and 5.50  \AA, respectively\cite{FeSeBulk}. Due to the lattice mismatch, the in-plane lattice parameter of single layer FeSe is enlarged to 3.901 \AA\cite{FeSeLayer}.  We used a slab model with 10 \AA\, vacuum space to calculate the total energy of single layer FeSe. The exchange correlation potential was described by the Perdew-Burke-Ernzerhof (PBE) type \cite{PBE1} of generalized gradient approximation (GGA). The kinetic energy cutoff for wavefunction was chosen to be 268.0 eV. A large rectangle supercell was used in genetic evolvement calculation, as shown by the green rectangle in Fig. \ref{Fig2}(a). Each supercell contains 32 Fe and 32 Se atoms. 4$\times$4$\times$6 and 4$\times$4$\times$1 Monkhorst-pack K-points settings were used in the reciprocal space integrations for FeSe bulk and single layer respectively.

We performed 30 generations of genetic evolution. In each generation, 30 populations including 14 combinations, 10 random structures, and 6 mutations were calculated. After the DFT calculations for one generation, eight best structures were selected to generate the population of the next generation. We set the total magnetic moment as 0, so that the system only searched for anti-ferromagnetic structuress. Fig. \ref{Fig2}(b) shows the calculation results of FeSe single layer. \texttt{MagGene} found the correct magnetic structure at the fourth generation. It is collinear antiferromagnetic as shown in the inner panel of Fig. \ref{Fig2}(b). The calculation results for FeSe bulk is shown in Fig. \ref{Fig2}(c).\texttt{MagGene} found the correct magnetic structure at the tenth generation. It is pair-checkerboard antiferromagnetic as shown in the inner panel of Fig. \ref{Fig2}(c). We would like to emphasize the importance of choosing a proper supercell. If the supercell is not compatable with the ground state, such as hexagonal supercell with rectangle ground state, then the true ground state could not be found out. That doesn't mean that the calculation results are totally useless. The state found by genetic algorithm probably has the siginature of the true ground state if the supercell is large enough. So the supercell should large enough, otherwise we need to perform genetic evolvement several times with different shapes of supercell.

\subsection{CrI$_3$ Single Layer}
CrI$_3$ single layer is the first two-dimensional intrinsic long-range magnet\cite{CrI3Monolayer}. It has rich magnetic properties\cite{CrI3Zheng,CrI3Yan}, and can be used in spin-tronic devices. The system contains magnetic Cr atoms and heavy I atoms. Each Cr has 3 $\mu$B magnetic moment. Due to the strong spin-orbit coupling in I atoms and the super-exchange involving both Cr and I atoms, the magnetic moment of Cr has large anisotropy, which eliminates the application of Mermin-Wagner theorem and conserves the long-range magnetic order. The system is very soft due to its very small Yang's modulus, therefore the lattice parameters can be largely tuned by lattice mismatch in experiment. And the different lattice parameters may lead to different magnetic structures\cite{CrI3Zheng}. 

  \begin{figure}
  \includegraphics[width=0.5\columnwidth]{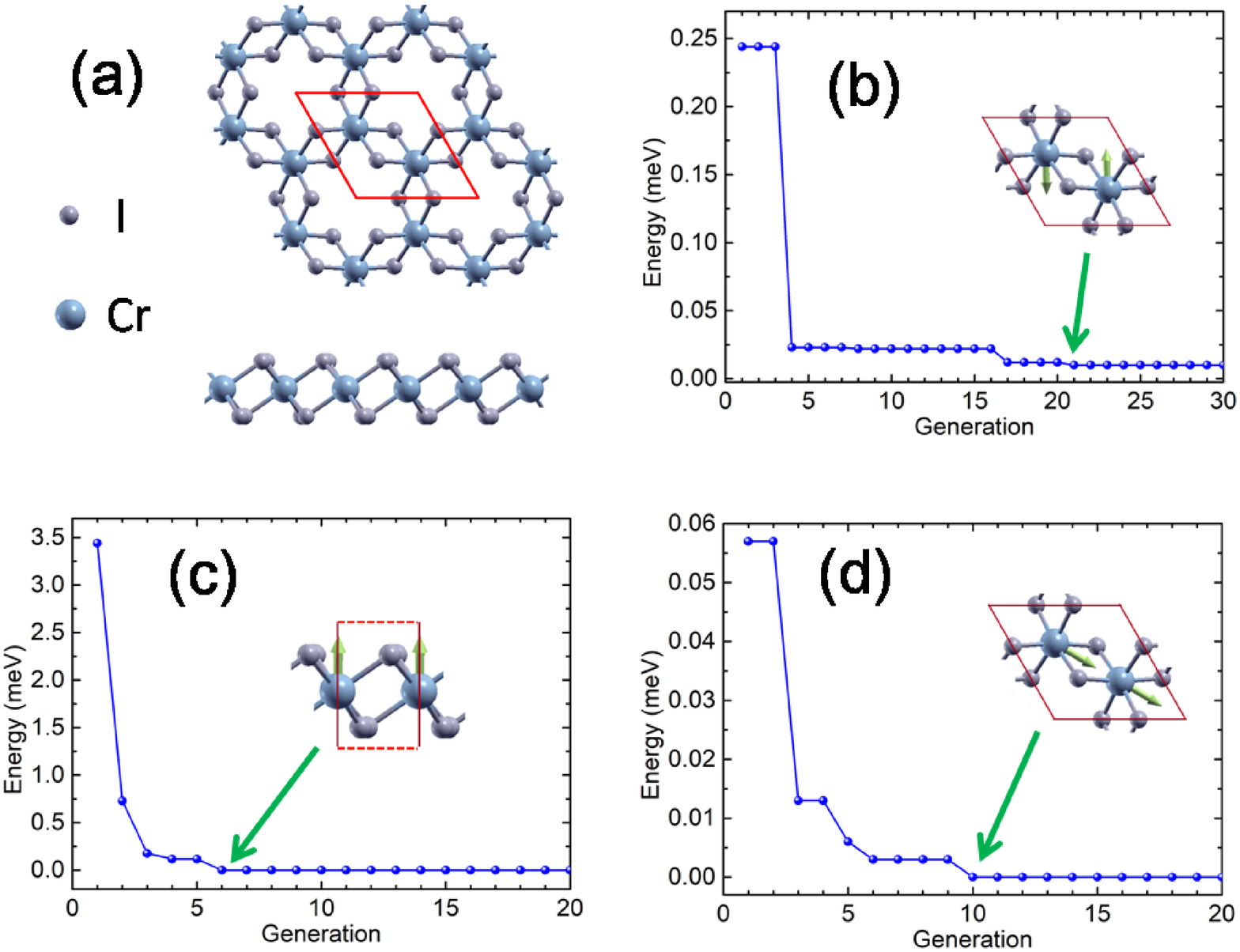}
  \caption{
  \label{Fig3}
   (a) Atomic structure of CrI$_3$ single layer. The primitive cell is shown by a red rhombus. Panels (b), (c) and (d) show the {\bf MagGene} calculation results for CrI$_3$ single layer with lattice parameters 6.3 \AA, 7.0 \AA, and 8.0 \AA\, respectively.
  }
  \end{figure}

As the second example, we predicted the magnetic structures of CrI$_3$ single layers with three different lattice parameters, which are 6.3 \AA, 7.0 \AA, and 8.0 \AA. The atomic structure of CrI$_3$ single layer is shown in Fig. \ref{Fig3}(a). It is a honeycomb lattice. In which, the large cyon and small gray balls  are Cr and I atoms respectively. The red rhombus shows the primitive cell. We used slab models with 15 \AA\, vacuum space to calculate their total energy. The exchange correlation potential was described by the generalized gradient approximation (GGA) in Perdew-Burke-Ernzerhof (PBE) type \cite{PBE1}. The kinetic energy cutoff for wavefunction was chosen to be 227.1 eV. A 8$\times$8$\times$1 Gamma centered K-mesh was used in the reciprocal space integration. The spin-orbit coupling correction was also included in our DFT calculations.

We performed 20-30 generations of genetic evolution. In each generation, 20 populations including 10 combinations, 3 random structures, and 7 mutations were calculated. Six best structures were selected to generate the next generation. Since the most stable magnetic orders include both ferromagnetic and anti-ferromagnetic orders for different lattice parameters, the total magnetic moments were not fixed. Fig. \ref{Fig3}(b) shows the calculation results of CrI$_3$ single layer with 6.3 \AA\, lattice parameter. At the fourth generation, \texttt{MagGene} found a magnetic structure close to in-plane antiferromagnetic configuration, and continued to optimize the structure until getting the standard in-plan antiferromagnetic order at the 21-th step, as shown in the inner panel of Fig. \ref{Fig3}(b). The calculation results of CrI$_3$ single layer with 7.0 \AA\, lattice parameter are shown in Fig. \ref{Fig3}(c). \texttt{MagGene} found the correct magnetic structure at the sixth generation. It is the out-plane ferromagnetic order. For the case of CrI$_3$ with 8.0 \AA\, lattice parameter, \texttt{MagGene} found the most stable magnetic structure at the eighth generation as shown in Fig. \ref{Fig3}(b). It is the in-plane ferromagnetic order.

\subsection{UO$_2$ Bulk}
Strong spin-orbit coupling may lead to noncollinear magnetic order. Due to the heavy relativistic, actinide atoms usually have very strong spin-orbit couplings. A typical example is UO$_2$ bulk. Expetimental studies found that the system may have peculiar 3-k noncollinear magnetic orders\cite{UO2Exp1,UO2Exp2,UO2Exp3}. Different magnetic orders have also been extensively studied in theory\cite{UO2Theory1,UO2LDApU,UO2LDApU2}. And the 3k type noncollinear magnetic order is usually not the most stable magnetic order in DFT calculations. In this example, we used SCAN functional\cite{SCAN1} to treat UO$_2$ bulk. It is a meta-GGA functional, and it is found significantly better than GGA and LDA functionals in many different systems\cite{SCAN2}.

\begin{figure}
  \includegraphics[width=0.5\columnwidth]{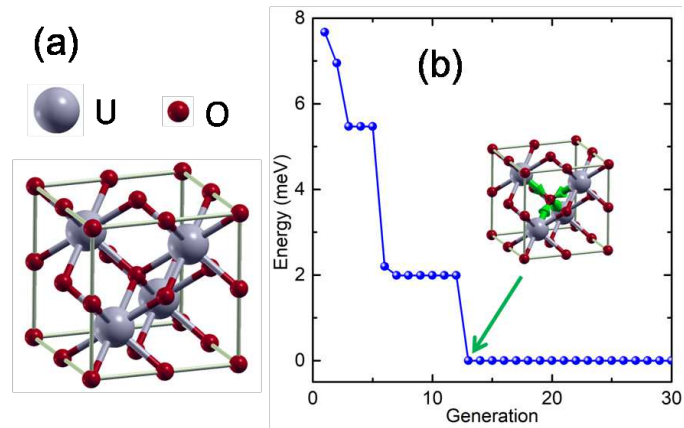}
  \caption{
  \label{Fig4}
   (a) Atomic structure of UO$_2$ bulk. The conventional cell is shown by a cyan cubic cage. Panels (b) shows the {\bf MagGene} calculation result.
  }
\end{figure}

The atomic structure of UO$_2$ bulk is shown in Fig. \ref{Fig4}(a). The large gray and small red balls are the U and O atoms respectively. The lattice parameters is chosen to be 5.469 \AA\, according to the experiment results. A cubic cell was used in our calculation. The kinetic energy cutoff for wavefunction was chosen to be 400 eV. A 4$\times$4$\times$4  Monkhorst-pack k-mesh was used in the reciprocal space integration. The spin-orbit coupling correction was included in our DFT calculations. Hubbard U correction was also considered in order to correctly describe the relatively strong electron correlations. The Hubbard U and J parameters were 3.46 eV and 0.30 eV respectively following a previous study\cite{UO2LDApU2}.

We performed 30 generations of genetic evolvement. In each generation, 30 populations including 17 crossovers, 3 random structures, and 10 mutations were calculated. Eight best structures were selected to generate the next generation. Since most of the previously proposed magnetic orders are antiferromagnetic orders, the total magnetic moments were fixed to zero in our calculation. Fig. \ref{Fig4}(b) shows the calculation results. \texttt{MagGene} found the correct magnetic structure at the thirteenth generation. It is 3-k noncollinear antiferromagnetic as shown in the inner panel of Fig. \ref{Fig4}(b).

\section{Conclusion}

In this communication we have introduced \texttt{MagGene}, a Fortran 90 code using genetic algorithm for predicting magnetic structures. It also contains techniques to enhance population diversity and emperical method to elimite unstable structures.
\texttt{MagGene} enables accurate and efficient predictions of both collinear and noncollinear magnetic structures. It is flexible, can be used in complex systems even when they contain multi kinds of magnetic atoms.


\section{Acknowledgments}

The research leading to these results has received fundings from Science Challenge Project under Grant No. TZ2016004, NSFC under Grant No. 11974056 and 11625415, NSFC-NSAF under Grant No. U1530258 and U1630248, and National Key R\&D Program of China under Grant No. 2017YFB0701502.

\end{document}